\documentclass[preprint,12pt]{elsarticle}

\usepackage{amssymb}
\usepackage{amsmath}

\usepackage{multirow}
\usepackage{algorithm}
\usepackage{algpseudocode}
\usepackage[table]{xcolor}
\usepackage{colortbl}  

\newtheorem{theorem}{Theorem}

\newtheorem{assumption}{Assumption}

\usepackage{makecell}

\begin{document}

\begin{frontmatter}



\title{FedMLAC: Mutual Learning Driven Heterogeneous Federated Audio Classification}




\author[0,1]{Jun Bai} 
\ead{jun.bai@mcgill.ca}

\author[2]{Rajib Rana} 
\ead{rajib.rana@unisq.edu.au}

\author[4]{Di Wu\corref{cor1}} 
\ead{di.wu@unisq.edu.au}

\author[3]{Youyang Qu} 
\ead{youyang.qu@data61.csiro.au}

\author[2]{Xiaohui Tao} 
\ead{xiaohui.tao@unisq.edu.au}

\author[4]{Ji Zhang} 
\ead{ji.Zhang@unisq.edu.au}

\author[5]{Carlos Busso} 
\ead{busso@cmu.edu}

\author[6]{Shivakumara Palaiahnakote} 
\ead{S.Palaiahnakote@salford.ac.uk}

\cortext[cor1]{Corresponding author}


\affiliation[0]{organization={School of Computer Science, McGill University},
            city={Montreal},
            country={Canada}}
            
\affiliation[1]{organization={Mila - Quebec AI Institute},
            city={Montreal},
            country={Canada}}

\affiliation[2]{organization={School of Mathematics, Physics and Computing, University of Southern Queensland},
            city={Springfield},
            country={Australia}}

\affiliation[3]{organization={Data61, CSIRO},
            city={Melbourne},
            country={Australia}}
            
\affiliation[4]{organization={School of Mathematics, Physics and Computing, University of Southern Queensland},
            city={Toowoomba},
            country={Australia}}

\affiliation[5]{organization={Language Technologies Institute, Carnegie Mellon University},
            city={Pittsburgh},
            country={USA}}
            
\affiliation[6]{organization={School of Science, Engineering and Environment, University of Salford},
            city={Manchester},
            country={UK}}

\begin{abstract}

Federated Learning (FL) offers a privacy-preserving framework for training audio classification (AC) models across decentralized clients without sharing raw data. However, Federated Audio Classification (FedAC) faces three major challenges: \textit{data heterogeneity}, \textit{model heterogeneity}, and \textit{data poisoning}, which degrade performance in real-world settings. While existing methods often address these issues separately, a unified and robust solution remains underexplored. We propose FedMLAC, a mutual learning-based FL framework that tackles all three challenges simultaneously. Each client maintains a personalized local AC model and a lightweight, globally shared \textit{Plug-in} model. These models interact via bidirectional knowledge distillation, enabling global knowledge sharing while adapting to local data distributions, thus addressing both data and model heterogeneity. To counter data poisoning, we introduce a Layer-wise Pruning Aggregation (LPA) strategy that filters anomalous \textit{Plug-in} updates based on parameter deviations during aggregation. Extensive experiments on four diverse audio classification benchmarks, including both speech and non-speech tasks, show that FedMLAC consistently outperforms state-of-the-art baselines in classification accuracy and robustness to noisy data.

\end{abstract}










\end{frontmatter}



\section{Introduction}
\label{sec:intro}

Audio classification (AC) enables key applications across diverse domains, including speech command recognition~\cite{wang2024watch}, emotion detection~\cite{tu2024adaptive}, and environmental sound classification~\cite{xu2024lightweight}. In practice, however, audio data are often generated and stored locally on edge devices such as smartphones, wearables, and IoT sensors, which raises concerns about data privacy and communication efficiency. These constraints motivate the adoption of Federated Learning (FL), which facilitates collaborative model training without sharing raw data~\cite{bai2022fedewa, bai2025non}, making it a suitable paradigm for real-world audio applications and leading to the emergence of Federated Audio Classification (FedAC)~\cite{tong2021federated}.
Despite its potential, FedAC encounters several critical challenges: data heterogeneity resulting from diverse speakers and devices~\cite{ahmad2023robust}, model heterogeneity due to varying client capabilities~\cite{bai2024unified}, and data poisoning arising from adversarial inputs~\cite{wu2024fedinverse}. These challenges compromise robustness and scalability, underscoring the need for a unified and resilient learning framework.

While interest in FedAC continues to grow across both speech and non-speech applications, 
existing research tends to address the challenges of data heterogeneity, model heterogeneity, and data poisoning in partial or disjointed ways. A substantial body of work focuses on mitigating heterogeneous non-IID (non-Independent and Identically Distributed) data distributions through strategies such as regularization techniques~\cite{grollmisch2025federated}, client selection~\cite{xu2025dynamic}, representation learning~\cite{LIN2023110396}, or robust aggregation mechanisms~\cite{chen2021federated, zhang2021distributed}. While these techniques improve global model utility under skewed data, they often assume a homogeneous model architecture shared across all clients, which is impractical in real-world deployments involving heterogeneous edge devices. To address system-level diversity, recent studies have explored flexible model configurations, including personalized layers~\cite{li2023fedtp}, decoupled training~\cite{kan2024parameter, xu2024application}, incremental learning~\cite{zhang2025federated}, or sparsified aggregation~\cite{ali2025efl}. However, such methods rarely account for the interaction between model and data heterogeneity, which can cause unstable convergence, degraded generalization, or fairness concerns when local models diverge significantly.

Another important challenge in FedAC is the presence of corrupted or unreliable local data, commonly referred to as data poisoning. Such corruption can arise not only from adversarial clients but also from natural sources, including label inconsistencies in subjective tasks (e.g., emotion recognition) and distorted audio recordings in uncontrolled environments. These issues can significantly impair model stability and convergence, especially when coupled with non-IID data or heterogeneous client models. To mitigate these effects, researchers have proposed defense strategies such as anomaly detection~\cite{senol2024privacy}, adversarial learning~\cite{almaiah2024novel}, client clustering~\cite{bhuyan2024unsupervised}, and model pruning~\cite{benazir2025privacy}. While effective in isolation, these methods are typically reactive and rarely consider the joint effects of data and model heterogeneity. Moreover, many introduce a trade-off between robustness and utility, often degrading performance on clean or low-resource clients. Thus, achieving resilience to data corruption without sacrificing personalization or fairness remains an open problem in FedAC.

To address the intertwined challenges of data heterogeneity, model heterogeneity, and data corruption in FedAC, we propose \textbf{FedMLAC}, a Mutual Learning-driven Federated Audio Classification framework designed to achieve both personalization and robustness. FedMLAC comprises two key components: \textbf{FedMLAC-Update} and \textbf{FedMLAC-Aggregation}. To handle data and model heterogeneity, each client maintains a personalized local AC model and a lightweight, globally shared \textit{Plug-in} model distributed by the server. These models engage in bidirectional mutual learning, where the \textit{Plug-in} model conveys global knowledge to the local model for generalization, while simultaneously absorbing client-specific information to enhance personalization. This design allows clients with diverse data distributions and model architectures to collaborate effectively without sharing raw data. To mitigate the effects of data corruption caused by adversarial behavior or noisy local inputs, FedMLAC incorporates a robust \textbf{Layer-wise Pruning Aggregation (LPA)} strategy in the server aggregation phase. Instead of naively averaging client updates, LPA selectively prunes anomalous parameter deviations at the layer level, thus filtering out unreliable \textit{Plug-in} model updates and enhancing the stability and integrity of global aggregation. Together, these two components enable FedMLAC to perform scalable, personalized, and communication-efficient AC learning across heterogeneous FL environments.

The main contributions of this work are summarized as follows:

\begin{itemize}
    \item We propose \textbf{FedMLAC}, a novel federated audio classification framework that simultaneously tackles the challenges of data heterogeneity, model heterogeneity, and data poisoning, which are rarely addressed in a unified manner by existing approaches.
    
    \item We develop a mutual-learning-based local update strategy and a robust layer-wise pruning aggregation (LPA) mechanism that facilitate personalized training while filtering adversarial updates and mitigating client divergence.
    
    \item We theoretically analyze the convergence of FedMLAC and empirically demonstrate its effectiveness under heterogeneous and adversarial audio learning conditions.

\end{itemize}

The remainder of this paper is structured as follows. Section~\ref{sec:relatedwork} presents a comprehensive review of related work in the field of FedAC. Section~\ref{sec:method} describes the proposed approach in detail, including its core components and technical implementation. Section~\ref{sec:convergence} provides a theoretical analysis of convergence. Section~\ref{sec:experiment} outlines the experimental setup and reports the results and analysis across multiple benchmark datasets. Finally, Section~\ref{sec:conclusion} concludes the paper by summarizing the key contributions and suggesting directions for future research.

\section{Related Work} \label{sec:relatedwork}

Recent work in Federated Audio Classification (FedAC) has improved privacy-preserving learning on distributed audio data, but most methods focus on isolated issues. We categorize related efforts around three core challenges: \textit{data heterogeneity}, \textit{model heterogeneity}, and \textit{data poisoning}.

\subsection{Addressing Data Heterogeneity}

Data heterogeneity is a core challenge in FedAC, as client audio data often differs in content, labels, and recording conditions. Existing works address this through regularization~\cite{chen2021federated, zhang2021distributed}, which incorporates proximal terms to reduce divergence between local and global models; client grouping~\cite{hsu2024cluster, lee2024language}, which organizes clients into semantically coherent subgroups; and adaptive training strategies such as dynamic participation scheduling~\cite{xu2025dynamic} and data-aware model adaptation~\cite{grollmisch2025federated} that improve robustness under non-stationary environments. To enhance generalization under domain shift, \cite{LIN2023110396} proposes contrastive representation learning with weak supervision, while \cite{leroy2019federated} uses local adaptation followed by federated averaging to manage ambient audio diversity. Communication-efficient designs~\cite{du2024communication, zhang2025federated} further support heterogeneity by decoupling model components, enabling sparse updates, and relaxing synchronization. While these methods improve FedAC performance under statistical diversity, most still assume consistent model structures and clean inputs, leaving open challenges in addressing both architectural and adversarial variability.

\subsection{Handling Model Heterogeneity}

Model heterogeneity in FedAC arises from varying client-side capabilities, including compute power, memory, and network bandwidth. To accommodate this, personalized modeling is a key strategy: split architectures~\cite{ali2025efl} allow clients to retain private output heads while sharing a global encoder, and task-specific personalization~\cite{li2023fedtp} assigns distinct heads for speaker and emotion classification. To reduce communication and computation costs, methods such as gradient sparsification~\cite{amiri2021federated}, encoder-head decoupling~\cite{zhu2022decoupled}, and bi-level optimization for flexible model configurations~\cite{kan2024parameter} have been proposed. Lightweight tuning schemes~\cite{feng2022federated} further adapt acoustic models with minimal overhead. In parallel, asynchronous and adaptive training schedules~\cite{zhang2025federated, du2024communication} support flexible client participation and balance communication load in streaming settings. While these approaches enable structural flexibility in FedAC, they often neglect robustness to noisy input conditions.

\subsection{Robustness Against Data Poisoning}

FedAC is vulnerable to data poisoning, as clients may submit mislabeled, corrupted, or adversarial audio. Existing methods aim to enhance robustness by detecting unreliable updates, filtering noise, or defending against malicious behavior. Anomaly-based defenses play a central role: Tsouvalas et al.~\cite{tsouvalas2024labeling} use confidence-based outlier detection, while~\cite{senol2024privacy} introduces a privacy-preserving verification mechanism via homomorphic encryption and statistical thresholds. Structural defenses modify model behavior to resist compromised updates, such as model pruning with trust scoring~\cite{jia2022federated} and selective gradient masking~\cite{benazir2025privacy}. Unsupervised methods provide complementary strategies under weak or unlabeled settings. Almaiah et al.~\cite{almaiah2024novel} utilize signal pre-processing and inter-client validation, while~\cite{bhuyan2024unsupervised} employ federated clustering to detect abnormal label patterns. Earlier work~\cite{tsouvalas2022federated} also introduces uncertainty-aware weighting to downscale noisy updates. While these defenses address key vulnerabilities in FedAC, they are rarely integrated with structural personalization or designed for data heterogeneous environments.

\subsection{Summary and Motivation}

As summarized in Table~\ref{tab:fAC_comparison}, existing FedAC methods have addressed challenges such as data heterogeneity~\cite{hsu2024cluster}, model diversity~\cite{zhu2022decoupled}, and data poisoning~\cite{tsouvalas2024labeling, almaiah2024novel}, but typically in isolation. Few works consider their interaction, despite the fact that in real-world audio applications, these issues often arise together. This fragmentation limits system robustness, fairness, and generalization—highlighting the need for more integrated approaches that can jointly address statistical, structural, and adversarial challenges in FedAC.

\begin{table*}[t]
    \centering
    \setlength{\tabcolsep}{3pt} 
    \fontsize{8}{10}\selectfont
    \caption{Systematic comparison of existing FedAC methods with respect to three core challenges: data heterogeneity, model heterogeneity, and data poisoning. Note: $\checkmark$ (Addressed); $\times$ (Not addressed).}
    \label{tab:fAC_comparison}
    \begin{tabular}{c|c|c|cc}
        \hline
        \textbf{Studies} & \textbf{Data Heterogeneity} & \textbf{Model Heterogeneity} & \textbf{Data Poisoning} \\
        \hline
        \cite{chen2021federated, zhang2021distributed, hsu2024cluster, LIN2023110396, leroy2019federated, grollmisch2025federated, xu2025dynamic, lee2024language}  & $\checkmark$ & $\times$  & $\times$ \\

        \cite{amiri2021federated, ali2025efl, li2023fedtp, kan2024parameter, feng2022federated}  & $\times$ & $\checkmark$  & $\times$ \\

         \cite{tsouvalas2022federated, tsouvalas2024labeling, senol2024privacy, jia2022federated, benazir2025privacy, almaiah2024novel, bhuyan2024unsupervised}  & $\checkmark$ & $\times$  & $\checkmark$ \\

        \cite{federatedtraining2023, zhu2022decoupled, du2024communication, zhang2025federated}  & $\checkmark$ & $\checkmark$  & $\times$ \\
        
        \rowcolor{gray!20}
        \textbf{FedMLAC (Ours)} & $\checkmark$ & $\checkmark$ &  $\checkmark$ \\
        \hline
    \end{tabular}
    
\end{table*}


\section{Methodology} \label{sec:method}

\subsection{Prelimineries} \label{sec:prelimineries}

\paragraph{\textbf{Problem Setup}}

Federated Audio Classification (FedAC) aims to train a global Audio Classification (AC) model $\omega^g$  collaboratively across multiple clients while keeping each client’s speech data decentralized to preserve privacy. Let there be \( N \) clients in the FedAC system, indexed by \( k \in \{1, 2, \ldots, N\} \). Each client \( k \) has a local AC dataset \( \mathcal{D}_k = \{(\mathbf{x}_{k,j}, \mathbf{y}_{k,j})\}_{j=1}^{n_k} \), where \( \mathbf{x}_{k,j} \) represents the input speech features and \( \mathbf{y}_{k,j} \) represents the corresponding label. The global objective of FedAC is to minimize the following loss function:

\begin{equation}
\label{eq:fl_obj}
_{\omega^g}^{min}\mathcal{L}(\omega^g)=\sum_{k=1}^{N}\frac{n_k}{n}\mathcal{L}_k(\omega^g),
\end{equation}
where $n=\sum_{k=1}^{N}{n_k}$, $n_k=|D_k|$, $\mathcal{L}_k(\omega^k)=\frac{1}{n_i} \sum_{j=1}^{n_i} \ell(\mathbf{\omega^k}; \mathbf{x}_{k,j}, \mathbf{y}_{k,j})$ denotes the average loss of the ${k^{th}}$ client model $\omega^k $ over $D_k$, and $\ell(\cdot)$ represents the training loss (e.g., cross-entropy (CE) loss for AC tasks).  

Each client updates its own local AC model \( \mathbf{\omega}^k \) by performing local training on its dataset \( \mathcal{D}_k \) using Stochastic Gradient Descent (SGD). The local update rule is:
\begin{equation}
\begin{aligned}
\omega ^k_{t + (i + 1)} = \omega ^k_{t + (i)} - {\eta _{t + (i)}}\nabla {\mathcal{L}_k}(\omega ^k_{t + (i)}),i = 0,1,...,E - 1,
\end{aligned}
\label{eq:sgd-update}
\end{equation}
where \( \eta \) is the learning rate, $\omega_t^k$ represents the $k^{th}$ client model in the $t^{th}$ communication round, and \( E \) denotes the local training epochs.

After local training, each client sends its updated model \( \mathbf{\omega}_{t+1}^{k} \) to the server, which performs aggregation to update the global model. The most commonly used aggregation method is Federated Averaging (FedAvg) \cite{fedavg}, defined as:
\begin{equation}
    \mathbf{\omega}_{t+1}^g = \sum_{k=1}^N \frac{n_k}{n} \mathbf{\omega}_{t+1}^{k}.
    \label{eq:avg_agg}
\end{equation}

The basic FedAC framework requires that each client's AC model \( \mathbf{\omega}^k \) maintains the same model architecture. Furthermore, the heterogeneity of distributed AC data leads to an inconsistency between the local learning objectives and the global FedAC training objective. These challenges significantly reduce the practicality and stability of existing FedAC frameworks.

\paragraph{\textbf{Mutual Learning}}

Mutual Learning is a collaborative training mechanism where two or more models interactively learn from each other by exchanging predictions during training. It enhances the learning capability of individual models by encouraging consistency and improving generalization \cite{zhang2018deep}. 

For simplicity, consider the case of two models, \( \mathbf{\omega}^i \) and \( \mathbf{\omega}^j \), learning from their respective datasets \( \mathcal{D}_i \) and \( \mathcal{D}_j \). The mutual learning objective for each model combines a task-specific loss and a consistency loss. The total loss for \( \mathbf{\omega}^i \) and \( \mathbf{\omega}^j \) is defined as:

\begin{equation}
\left\{
\begin{aligned}
\mathcal{L}_i &= \mathcal{L}_{\text{task}}(\mathbf{\omega}^i; \mathcal{D}_i) + \lambda \mathcal{L}_{\text{KL}}(\mathit{p}_i, \mathit{p}_j) \\
\mathcal{L}_j &= \mathcal{L}_{\text{task}}(\mathbf{\omega}^j; \mathcal{D}_j) + \lambda \mathcal{L}_{\text{KL}}(\mathit{p}_j, \mathit{p}_i)
\end{aligned}
\right.
\end{equation}
where \( \mathcal{L}_{\text{task}}(\mathbf{\omega}^i; \mathcal{D}_i) \) is the task-specific loss (e.g., CE loss) for model \( \omega^i \) on dataset \( \mathcal{D}_i \).
\( \mathcal{L}_{\text{KL}}(\mathit{p}_i, \mathit{p}_j) \) is the Kullback-Leibler (KL) divergence between the output distributions \( \mathit{p}_i \) and \( \mathit{p}_j \) of models \( \mathbf{\omega}^i \) and \( \mathbf{\omega}^j \), defined as:
\begin{equation}
    \mathcal{L}_{\text{KL}}(\mathit{p}_i, \mathit{p}_j) = \sum_{c} \mathit{p}_i(c) \log \frac{\mathit{p}_i(c)}{\mathit{p}_j(c)},
\end{equation}
where \( \mathit{p}_i(c) \) represents the probability of class \( c \) predicted by model \( \mathbf{\omega}^i \). \( \lambda \) is a hyperparameter balancing the task loss and consistency loss. The models \( \mathbf{\omega}^i \) and \( \mathbf{\omega}^j \) are trained iteratively to minimize their respective losses \( \mathcal{L}_i \) and \( \mathcal{L}_j \), leading to mutual enhancement of their predictive performance. This learning paradigm is particularly beneficial in scenarios with model heterogeneity.

\subsection{FedMLAC}

Our proposed method consists of two main components: \textbf{FedMLAC-Update} and \textbf{FedMLAC-Aggregation}. The first component, FedMLAC-Update, takes place on the federated clients and aims to update both the local client AC models and the global \textit{Plug-in} AC model through a mutual learning mechanism. This process facilitates the transfer of global federated knowledge to the client AC models while simultaneously updating the global \textit{Plug-in} AC model with new local knowledge.

\begin{figure*}[tb]
\centering
\setlength{\abovecaptionskip}{0cm}
\includegraphics[width=1.0\linewidth,scale=1.0]{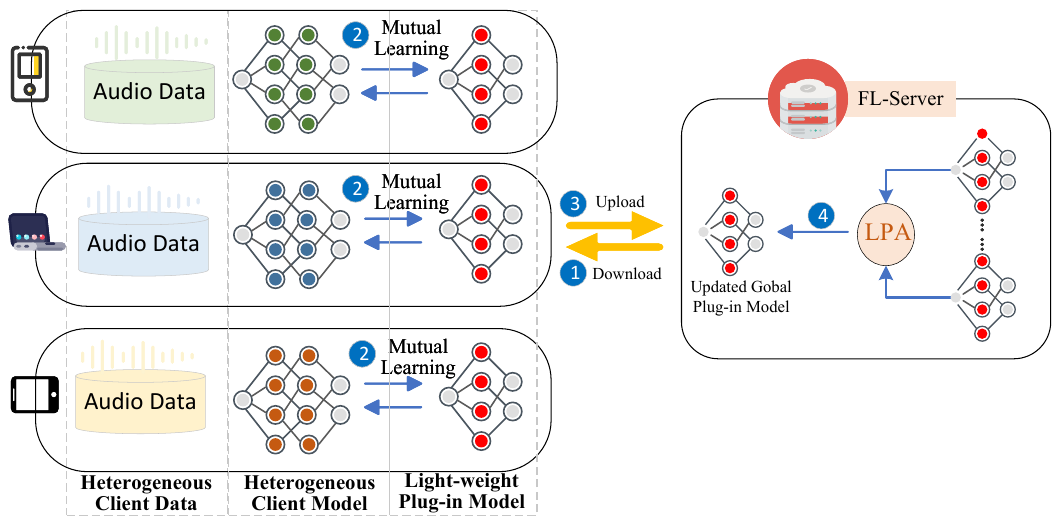}
\caption{Overview of the proposed FedMLAC framework. The framework consists of four key steps: \textcircled{1} each client downloads the globally shared lightweight \textit{Plug-in} model from the server; \textcircled{2} local mutual learning is performed between the personalized client model and the \textit{Plug-in} model to enable bidirectional knowledge transfer; \textcircled{3} the updated \textit{Plug-in} models are uploaded to the server; \textcircled{4} the server aggregates these models using the proposed Layer-wise Pruning Aggregation (LPA) strategy to form a new global \textit{Plug-in} model. This iterative process enables continuous learning across heterogeneous clients while improving robustness to noisy or non-IID data.}
\label{fig:fedmlAC-overview}
\end{figure*}

The second component, FedMLAC-Aggregation, operates on the server side. It aggregates the updated \textit{Plug-in} AC models uploaded by the clients, resulting in an updated global \textit{Plug-in} AC model. This updated global AC model serves as an auxiliary tool for the subsequent round of client AC model updates, as illustrated in Figure~\ref{fig:fedmlAC-overview}.

\subsubsection{FedMLAC Update}

\begin{algorithm}[t]
    \caption{\textit{FedMLAC-Update}}
    \label{alg:fedmlAC-update}
    \begin{algorithmic}
        \State \textbf{Input:} Server-shared global \textit{Plug-in} model $\theta_t^k$ and client model $\omega_{t}^k$
        \State \textbf{Output:} Updated $\theta_{t+1}^k$ and $\omega_{t+1}^k$ 
    \end{algorithmic}
    \begin{algorithmic}[1]
        \Procedure{\textnormal{FedMLAC-Update($\theta_t^k$,$k$,$t$):}}{}
            \State \hfill \(\triangleright\) Run on the client $k$
            \For {each epoch $i$ from $1$ to $E$}
                \For {each batch $\mathcal{B} \in \mathcal{D}_k$}
                    \State Compute loss $\mathcal{L}^{k}(\mathcal{B}; \omega^k_{t}, \theta_t^k)$ using Equation (\ref{eq:client_obj}) 
                    \State \hfill \(\triangleright\) Client model training loss     
                    \State Update client model parameters using SGD:
                    \State  $
                        \omega^k_{t+(i+1)} \leftarrow \omega^k_{t+(i)} - \eta \nabla \mathcal{L}^k(\mathcal{B}; \omega^k_{t+(i)}, \theta^k_{t+(i)})
                     $
                    \State \hfill \(\triangleright\) $\eta$ is the learning rate.

                    \State Compute loss $\mathcal{L}^{g}(\mathcal{B};\omega^k_t, \theta^k_t)$ using Equation (\ref{eq:gkd}) 
                    \State \hfill \(\triangleright\) \textit{Plug-in} model training loss   \State Update \textit{Plug-in} model parameters using SGD:
                    \State  $
                        \theta^k_{t+(i+1)} \leftarrow \theta^k_{t+(i)} - \eta \nabla \mathcal{L}^g(\mathcal{B}; \omega^k_{t+(i)}, \theta^k_{t+(i)})
                     $
                    
                \EndFor
            \EndFor
           
            \State return $\theta ^k_{t + 1}$ 
        \EndProcedure
    \end{algorithmic}
\end{algorithm}

Unlike traditional FedAC frameworks, our FedMLAC framework allows each client to have its own personalized AC model, making the framework more aligned with real-world applications. Additionally, each client receives a global \textit{Plug-in} AC model from the server, which participates in the training process alongside the local AC model. The specific training objectives for these two AC models are described below.

\paragraph{\textbf{Local AC Model Update}}

At the beginning of each federated communication round, the local AC model $\omega^k_t$ undergoes training and parameter updates on the private dataset. Consequently, its training objective must include an AC task-specific loss function, such as the commonly used CE loss, which is defined as follows:
\begin{equation}
\label{eq:lce}
\mathcal{L}^{k}_{CE}((x,y);\omega^k_t)=CE(F_k(x;\omega^k_t),y),
\end{equation}
where $(x,y) \in D_k$, and $F_k$ represents the $k^{th}$ client AC model. 

In addition to the AC task-specific loss, to obtain a more adaptable and robust local AC model, each client participating in FL should acquire knowledge about the AC data distributions of other clients. This can be achieved by treating the global \textit{Plug-in} AC model  $\theta^k_t$ as a teacher and the local AC model as a student, enabling the local AC model to effectively learn global knowledge. Specifically, a KD loss is added to the local AC model’s training objective, defined as:

\begin{equation}
\label{eq:lkd}
\mathcal{L}^{k}_{KL}(x;\omega^k_t, \theta^k_t)= KL\left(F_g(x;\theta^k_t), F_k(x;\omega^k_t))\right),
\end{equation}
wherein $KL(\cdot,\cdot)$ denotes the KL divergence loss. By combining Equations. (\ref{eq:lce}) and (\ref{eq:lkd}), we reformulate the client AC model training objective function in the $(t+1)^{th}$ communication round as:
\begin{equation}
\label{eq:client_obj}
_{\omega^k_{t+1}}^{min}\mathcal{L}^{k}((x,y); \omega^k_{t}, \theta_t^k)=\alpha\mathcal{L}^{k}_{CE}+(1-\alpha)\mathcal{L}_{KL}^{k},
\end{equation}
where \( \alpha \) is a balancing parameter that controls the relative weight of the two loss terms.

\paragraph{\textbf{Global \textit{Plug-in} Model Update}}

For the global \textit{Plug-in} model, in addition to serving as a carrier of global knowledge, it also participates in training on each client to learn the local AC data knowledge of that client. This enables the \textit{Plug-in}  model to update its existing knowledge based on local client data. In this case, the local client AC model acts as the teacher, and the \textit{Plug-in} model serves as the student. Consequently, the training objective for updating the \textit{Plug-in}  model includes a knowledge distillation loss, defined as:

\begin{equation}
\label{eq:gkd}
\mathcal{L}^{g}(x;\omega^k_t, \theta^k_t)= KL\left(F_k(x;\omega^k_t), F_g(x;\theta^k_t))\right).
\end{equation}

The loss of Equation (\ref{eq:gkd}) ensures that the \textit{Plug-in} model effectively integrates client-specific AC data knowledge, enhancing its adaptability and robustness for global FL. The entire FedMLAC-Update process is detailed in Algorithm~\ref{alg:fedmlAC-update}.

\subsubsection{FedMLAC Aggregation}

Once the global \textit{Plug-in} model on each client completes its local knowledge update, it is uploaded to the server for aggregation to generate a new global \textit{Plug-in}  model. In traditional federated aggregation (e.g., Equation (\ref{eq:avg_agg})), all client-uploaded \textit{Plug-in} models $\{\theta_{t+1}^{k}\}_{k\in S_t}$ are directly combined using data-size weighted averaging. However, under non-IID data distributions or adversarial conditions (e.g., noisy inputs, label-flipping, or poisoned data), some local \textit{Plug-in}  models may introduce substantial deviations from the true global optimum. These deviations manifest as abnormally large or small parameters in specific layers, thereby biasing the global \textit{Plug-in} model when such updates are indiscriminately aggregated.

To address this, we introduce a robust layer-wise aggregation strategy named \textbf{Layer-wise Pruning Aggregation (LPA)}, which aims to eliminate extreme deviations by pruning abnormal client updates at each individual layer before aggregation. This strategy improves robustness against both statistical heterogeneity and malicious updates by identifying outlier layers that exhibit inconsistent behavior relative to the majority.

\paragraph{\textbf{Layer-wise Similarity Computation}}

Let $\theta_{t+1}^{k} = \{\theta_{t+1}^{k,l}\}_{l=1}^{L}$ denote the $k^{th}$ client’s \textit{Plug-in} AC model at round $t+1$, where $\theta_{t+1}^{k,l}$ is the parameter vector of the $l^{th}$ layer. For each layer $l$, we define the average model across clients as:
\begin{equation}
\label{eq:lpa_avg}
\bar{\theta}_{t+1}^{l} = \frac{1}{|S_t|} \sum_{k \in S_t} \theta_{t+1}^{k,l}.
\end{equation}

We then measure the deviation of each \textit{Plug-in} model’s layer parameters from the average using the $\ell_2$ norm:
\begin{equation}
\label{eq:lpa_deviation}
\delta^{k,l} = \|\theta_{t+1}^{k,l} - \bar{\theta}_{t+1}^{l} \|_2.
\end{equation}

This value reflects how much \textit{Plug-in} model $k$'s layer-$l$ parameters deviate from the average behavior of the cohort. A high deviation may indicate either statistical heterogeneity or an adversarial anomaly.

\begin{algorithm}[t]
    \caption{\textit{FedMLAC-Aggregation}}
    \label{alg:fedmlAC-aggregation}
    \begin{algorithmic}
        \State \textbf{Input:} Client-uploaded \textit{Plug-in} models $\{\theta_{t+1}^k\}_{k \in S_t}$ 
        \State \textbf{Output:} Aggregated global \textit{Plug-in} model $\theta_{t+1}^g$ 
    \end{algorithmic}
    \begin{algorithmic}[1]
        \Procedure{\textnormal{FedMLAC-Aggregation($\{\theta_{t+1}^k\}_{k \in S_t}$)}}{}
            \State \hfill \(\triangleright\) Run on the server
            \For {each model $k \in  S_t$}
            \For {each model layer $l$ from $1$ to $L$}
                \State Compute  $\bar{\theta}_{t+1}^{l}$ using Equation (\ref{eq:lpa_avg})
                \State \hfill \(\triangleright\)  average model layer across $\{\theta_{t+1}^k\}_{k \in S_t}$)

                \State Compute  $\delta^{k,l}$ using Equation (\ref{eq:lpa_deviation})
                \State \hfill \(\triangleright\)  $l^{th}$  model layer deviation

                \State Collect layer deviations: $\bar{\mathcal{T}}_l \leftarrow \delta^{k,l}$
                         
            \EndFor
         \EndFor

        \State Sort and prune $\bar{\mathcal{T}}_l$ as $\mathcal{T}_l$ according to Equation (\ref{eq:trusted_set})
          \State \hfill \(\triangleright\)  prune high-deviation layers
        
        \State Compute   $\theta_{t+1}^{g,l} = \sum_{k \in \mathcal{T}_l} \frac{n_k}{n_{\mathcal{T}_l}} \theta_{t+1}^{k,l}$
        \State \hfill \(\triangleright\)  $l^{th}$  model layer-wise pruning aggregation

        \State Form the new global \textit{Plug-in} model $\theta_{t+1}^{g} = \{\theta_{t+1}^{g,l}\}_{l=1}^L$
               
        \EndProcedure
    \end{algorithmic}
\end{algorithm}

\paragraph{\textbf{Pruning High-Deviation Layers}}

To mitigate the influence of outlier updates, we prune the \textit{Plug-in} models that exhibit extreme deviations at each layer. Specifically, for each layer $l$, we first compute the deviation scores $\delta^{k,l}$ for all clients $k \in S_t$ using Equation~(\ref{eq:lpa_deviation}). These scores are then sorted in ascending order, and both the top-$\lfloor v_h \cdot |S_t| \rfloor$ and bottom-$\lfloor v_l \cdot |S_t| \rfloor$ clients are removed from consideration. The remaining clients form the trusted subset $\mathcal{T}_l$ used for layer-$l$ aggregation, formally defined as:
\begin{equation}
\label{eq:trusted_set}
\mathcal{T}_l = \left\{ k \in S_t \;\middle|\; \delta^{k,l} \in \text{Middle}\left( \left\{ \delta^{k,l} \right\}_{k \in S_t}, v_l, v_h \right) \right\},
\end{equation}
where the operator $\text{Middle}(\cdot, v_l, v_h)$ returns the central subset of deviations after removing the largest $\lfloor v_h \cdot |S_t| \rfloor$ and smallest $\lfloor v_l \cdot |S_t| \rfloor$ elements. Accordingly, the size of the trusted set is:
\begin{equation}
\label{eq:trusted_size}
|\mathcal{T}_l| = |S_t| - \lfloor v_h \cdot |S_t| \rfloor - \lfloor v_l \cdot |S_t| \rfloor.
\end{equation}

This pruning strategy helps reduce the impact of statistical heterogeneity or adversarial perturbations by excluding model updates with extreme deviations prior to aggregation.


The pruned, layer-wise aggregation is then performed as:
\begin{equation}
\label{eq:lpa_final}
\theta_{t+1}^{g,l} = \sum_{k \in \mathcal{T}_l} \frac{n_k}{n_{\mathcal{T}_l}} \theta_{t+1}^{k,l},
\end{equation}
where $n_k$ is the dataset size of client $k$ and $n_{\mathcal{T}_l} = \sum_{k \in \mathcal{T}_l} n_k$ is the total data size of the trusted clients for layer $l$. The final global \textit{Plug-in} model $\theta_{t+1}^{g}$ is then reconstructed by concatenating all the aggregated layers: $\theta_{t+1}^{g} = \{\theta_{t+1}^{g,l}\}_{l=1}^L$. The specific implementation procedure of FedMLAC-Aggregation is provided in Algorithm \ref{alg:fedmlAC-aggregation}. 


\subsection{Framework Insights}

FedMLAC embodies a unified design that integrates personalization, scalability, and robustness in FedAC by decoupling global knowledge aggregation from personalized learning through a collaborative \textit{Plug-in} mechanism. This separation supports model heterogeneity and allows clients to retain autonomy over model structures and local objectives, which is essential in diverse acoustic scenarios with varying tasks, devices, and environments. By framing mutual learning as bidirectional knowledge distillation, FedMLAC balances generalization and adaptation within each training round, mitigating the trade-off between convergence stability and personalization that often hinders traditional FL pipelines. The LPA component further enhances resilience to noisy or corrupted updates, which is critical for audio data affected by ambiguous labels, device distortions, or inconsistent recording quality.

\section{Convergence Analysis of FedMLAC}
\label{sec:convergence}

This section presents a convergence analysis of the FedMLAC framework, focusing on the global \textit{Plug-in} model's convergence under the challenges of data heterogeneity, model heterogeneity, and data poisoning. 
We outline the theoretical assumptions, key analytical
steps, and the main convergence result.

\subsection{Theoretical Assumptions}

To facilitate the convergence analysis, we adopt the following standard assumptions, commonly used in FL literature:

\begin{assumption}[Smoothness]
\label{assump:smoothness}
The local loss functions $\mathcal{L}^k(\omega)$ and $\mathcal{L}^g(\theta)$ are $L$-smooth, i.e., for all $\omega, \omega', \theta, \theta'$:
\begin{equation}
\|\nabla \mathcal{L}^k(\omega) - \nabla \mathcal{L}^k(\omega')\| \leq L \|\omega - \omega'\|, \quad \|\nabla \mathcal{L}^g(\theta) - \nabla \mathcal{L}^g(\theta')\| \leq L \|\theta - \theta'\|.
\end{equation}
\end{assumption}

\begin{assumption}[Bounded Gradients]
\label{assump:bounded_gradients}
The gradients of the local loss $\mathcal{L}^k(\omega)$ and the \textit{Plug-in} model loss $\mathcal{L}^g(\theta)$ are bounded:
\begin{equation}
\|\nabla \mathcal{L}^k(\omega)\|^2 \leq G^2, \quad \|\nabla \mathcal{L}^g(\theta)\|^2 \leq G^2.
\end{equation}
\end{assumption}

\begin{assumption}[Bounded Data Heterogeneity]
\label{assump:heterogeneity}
The local loss gradients exhibit bounded heterogeneity with respect to the global loss:
\begin{equation}
\mathbb{E}[\|\nabla \mathcal{L}^k(\theta) - \nabla \mathcal{L}(\theta)\|^2] \leq \sigma^2,
\end{equation}
where $\sigma^2$ quantifies the degree of non-IID data across clients.
\end{assumption}

\begin{assumption}[Bounded Stochastic Gradient Variance]
\label{assump:variance}
The stochastic gradients computed on mini-batches $\mathcal{B} \subset \mathcal{D}_k$ are unbiased with bounded variance:
\begin{equation}
\mathbb{E}[\nabla \mathcal{L}^k(\omega; \mathcal{B})] = \nabla \mathcal{L}^k(\omega), \quad \mathbb{E}[\|\nabla \mathcal{L}^k(\omega; \mathcal{B}) - \nabla \mathcal{L}^k(\omega)\|^2] \leq \sigma_g^2,
\end{equation}
\begin{equation}
\mathbb{E}[\nabla \mathcal{L}^g(\theta; \mathcal{B})] = \nabla \mathcal{L}^g(\theta), \quad \mathbb{E}[\|\nabla \mathcal{L}^g(\theta; \mathcal{B}) - \nabla \mathcal{L}^g(\theta)\|^2] \leq \sigma_g^2.
\end{equation}
\end{assumption}

\begin{assumption}[LPA Effectiveness]
\label{assump:lpa}
The LPA strategy ensures that the parameters of the trusted subset $\mathcal{T}_l$ for each layer $l$ have bounded deviation from the average:
\begin{equation}
\mathbb{E}[\|\theta_{t+1}^{k,l} - \bar{\theta}_{t+1}^l\|^2] \leq \delta^2, \quad k \in \mathcal{T}_l,
\end{equation}
where $\bar{\theta}_{t+1}^l = \frac{1}{|S_t|} \sum_{k \in S_t} \theta_{t+1}^{k,l}$, and $\delta^2$ is the upper bound on the deviation after pruning.
\end{assumption}

\begin{assumption}[Bounded Loss]
\label{assump:bounded_loss}
The global loss $\mathcal{L}(\theta)$ is bounded below:
\begin{equation}
\mathcal{L}(\theta) \geq \mathcal{L}^* > -\infty.
\end{equation}
\end{assumption}

These assumptions ensure that the loss functions are well-behaved, gradients are manageable, and the LPA strategy effectively mitigates outliers due to data poisoning or heterogeneity. While we assume non-convexity (common for deep learning models), the smoothness and boundedness assumptions provide sufficient structure for convergence analysis.

\subsection{Convergence Analysis}

We analyze the convergence of $\theta^g$ to a stationary point of $\mathcal{L}(\theta^g)$ in three steps:  
(1) bounding local update progress,  
(2) analyzing LPA’s effect on aggregation, and  
(3) deriving the global convergence rate.  

Clients update $\omega^k$ and $\theta^k$ via SGD on Equations~(\ref{eq:client_obj}) and~(\ref{eq:gkd}). For $\omega^k$:
\begin{equation}
    \omega^k_{t+(i+1)} = \omega^k_{t+(i)} - \eta \nabla \mathcal{L}^k(\mathcal{B}; \omega^k_{t+(i)}, \theta^k_{t+(i)}),
    \label{eq:4}
\end{equation}
where $i = 0, \dots, E-1$.  
By $L$-smoothness (Assumption~\ref{assump:smoothness}), the expected loss decrease is:
\begin{equation}
    \mathbb{E}[\mathcal{L}^k(\omega^k_{t+(i+1)})] \leq \mathcal{L}^k(\omega^k_{t+(i)}) - \eta \|\nabla \mathcal{L}^k(\omega^k_{t+(i)})\|^2 + \frac{L\eta^2}{2}(G^2 + \sigma_g^2).
    \label{eq:5}
\end{equation}
Summing over $E$ epochs:


\begin{equation}
    \mathbb{E}[\mathcal{L}^k(\omega^k_{t+1})] \leq \mathcal{L}^k(\omega^k_{t}) - \eta \sum_{i=0}^{E-1} \|\nabla \mathcal{L}^k(\omega^k_{t+(i)})\|^2 + \frac{L\eta^2 E}{2}(G^2 + \sigma_g^2).
    \label{eq:6}
\end{equation}
Similarly, for $\theta^k$:
\begin{equation}
    \mathbb{E}[\mathcal{L}^g(\theta^k_{t+1})] \leq \mathcal{L}^g(\theta^k_t) - \eta \sum_{i=0}^{E-1} \|\nabla \mathcal{L}^g(\theta^k_{t+(i)})\|^2 + \frac{L\eta^2 E}{2}(G^2 + \sigma_g^2).
    \label{eq:7}
\end{equation}
Equation~(\ref{eq:7}) reflects the \textit{Plug-in} model’s progress, driven by mutual learning with $\omega^k$.  
The KL divergence term in Equation~(\ref{eq:client_obj}) regularizes $\omega^k$ toward global knowledge, reducing the impact of data heterogeneity ($\sigma^2$).

In FedMLAC-Aggregation, LPA computes per-layer averages $\bar{\theta}^l_{t+1}$, prunes outliers based on deviations $\delta^{k,l} = \|\theta^{k,l}_{t+1} - \bar{\theta}^l_{t+1}\|_2$, and aggregates:
\begin{equation}
    \theta^{g,l}_{t+1} = \sum_{k \in \mathcal{T}_l} \frac{n_k}{n_{\mathcal{T}_l}} \theta^{k,l}_{t+1}.
    \label{eq:8}
\end{equation}
Equation~(\ref{eq:8}) forms the global model by weighting trusted client updates, with  
$\mathbb{E}[\|\theta^{k,l}_{t+1} - \bar{\theta}^l_{t+1}\|^2] \leq \delta^2$ (Assumption~\ref{assump:lpa}).  
The global loss decrease is:
\begin{equation}
    \mathbb{E}[\mathcal{L}(\theta^g_{t+1})] \leq \mathcal{L}(\theta^g_t) - \eta \sum_{l=1}^L \sum_{k \in \mathcal{T}_l} \frac{n_k}{n_{\mathcal{T}_l}} \sum_{i=0}^{E-1} \|\nabla \mathcal{L}^g(\theta^k_{t+(i)})\|^2 + \frac{L\eta^2 E}{2}(G^2 + \sigma_g^2 + \sigma^2 + \delta^2).
    \label{eq:9}
\end{equation}
\subsection{Convergence Theorem}

\begin{theorem} \label{thm:convergence} Consider the training dynamics of the global \textit{Plug-in} model $\theta^g$ in FedMLAC under Assumptions~\ref{assump:smoothness}–\ref{assump:bounded_loss}. If the learning rate satisfies $\eta < \frac{1}{L}$, then after $T$ communication rounds, the average squared norm of the gradient of the global loss function $\mathcal{L}(\theta^g)$ satisfies:
\begin{equation}
    \frac{1}{T} \sum_{t=0}^{T-1} \mathbb{E}\left[\|\nabla \mathcal{L}(\theta^g_t)\|^2\right] 
    \leq \frac{\mathcal{L}(\theta^g_0) - \mathcal{L}^*}{\eta T E} + \frac{L \eta E}{2} (G^2 + \sigma_g^2 + \sigma^2 + \delta^2).
    \label{eq:10}
\end{equation}
\end{theorem}

Equation~(\ref{eq:10}) establishes that the global \textit{Plug-in} model converges to a stationary point at a rate of $\mathcal{O}(1/\sqrt{T})$ (with $\eta \propto 1/\sqrt{T}$). The bound captures error contributions from gradient magnitude ($G^2$), local stochastic variance ($\sigma_g^2$), data heterogeneity ($\sigma^2$), and deviations induced by LPA pruning ($\delta^2$). 


\subsection{Discussion}

The mutual learning mechanism, via KL divergence in Equations~(\ref{eq:client_obj}) and ~(\ref{eq:gkd}), aligns local models with global knowledge, reducing the heterogeneity term $\sigma^2$. This ensures that local updates contribute effectively to the global objective, even with non-IID data. LPA’s pruning strategy bounds $\delta^2$ by excluding anomalous updates, enhancing robustness against data poisoning. By focusing on trusted subsets, LPA mitigates the impact of corrupted or outlier client updates, ensuring stable aggregation.

The $\mathcal{O}(1/\sqrt{T})$ rate is standard for non-convex FL. The error terms reflect challenges inherent to distributed optimization. Limitations include the non-convexity assumption and fixed LPA parameters $v_h$, $v_l$. Future work could explore adaptive pruning or convex settings to further optimize convergence.

\section{Experiments} \label{sec:experiment}

In this section, we conduct comprehensive experiments to evaluate the effectiveness of the proposed FedMLAC framework. Our evaluation spans four benchmark datasets covering both speech and non-speech audio classification tasks. We assess FedMLAC in terms of classification accuracy, robustness to noisy and heterogeneous data, and personalization capability, comparing it against four representative state-of-the-art FL baselines.

\subsection{Datasets and Models}

\begin{table*}[t]
    \centering
    \setlength{\tabcolsep}{3pt} 
    \fontsize{8}{10}\selectfont
    \caption{Summary of federated settings across benchmark datasets, including audio classification tasks, partition strategies, and data statistics.}
    \begin{tabular}{c|c|c|c|c|c}
    \hline
        \textbf{Dataset} & \textbf{AC Task} & \textbf{Partition Strategy} & \textbf{\#Clients} & \textbf{\#Samples} & \textbf{\#Classes} \\ \hline
        GSC & Keyword Spotting & Speaker ID & 2,618 & 105,829 & 35 \\
        IEMOCAP & Emotion Recognition & Actor ID & 10 & 2,943 & 4 \\
        CREMA-D & Emotion Recognition & Speaker ID & 91 & 4,798 & 4 \\
        Urban Sound & Sound Event Detection & Dirichlet Distribution & 50 & 8,732 & 10 \\ \hline
    \end{tabular}
    \label{tab:fedac_datasets}
\end{table*}

\textbf{Datasets.} 
We evaluate FedMLAC on four publicly available datasets widely adopted in federated audio classification (FedAC), as summarized in Table~\ref{tab:fedac_datasets}. 
(1) Google Speech Commands 
(GSC)~\cite{warden2018speech} is a large-scale benchmark for keyword spotting, comprising over 100,000 short utterances covering 35 command words.
(2) Urban Sound~\cite{salamon2014dataset} includes environmental audio samples spanning 10 everyday sound event classes, such as sirens, drilling, and dog barking.
(3) IEMOCAP~\cite{busso2008iemocap} is a multimodal dataset for emotion recognition. In this work, we focus on its audio-only component, selecting four commonly used emotions from the improvised sessions.
(4) CREMA-D~\cite{cao2014crema} features speech clips performed by actors under emotional prompts. We consider the audio channel and select four target emotions consistent with prior studies.
To ensure fair comparison across datasets, we apply consistent preprocessing strategies including silence trimming, segmentation, and acoustic feature extraction (e.g., Mel spectrograms or pretrained Autoregressive Predictive Coding (APC) embeddings), following the protocol proposed in~\cite{zhang2023fedaudio}.

 \textbf{Data Partition.}  
To simulate realistic client distributions in federated settings, we adopt dataset-specific non-IID partitioning strategies. For the GSC and CREMA-D datasets, we assign samples to clients based on speaker identity, ensuring that each client contains utterances from a single speaker. This setup reflects the naturally occurring speaker-level data separation commonly found in speech-based federated systems.
In the case of IEMOCAP, we divide data by actor identity, with each client corresponding to the recordings of one actor. This preserves speaker consistency within clients while exposing variation across clients.
For the Urban Sound dataset, we follow a label distribution skewing strategy using the Dirichlet allocation method, as commonly adopted in FL studies~\cite{chen2020fedbe}. Specifically, we vary the Dirichlet concentration parameter $\alpha \in \{0.1, 0.3, 0.5, 1.0\}$ to control the degree of data heterogeneity. Smaller values of $\alpha$ lead to greater data heterogeneity across clients, whereas larger values result in more balanced and uniform data distributions.

\textbf{AC Models.} To support AC in both homogeneous and heterogeneous FL settings, we construct a suite of Convolutional Recurrent Neural Network (CRNN) architectures with varying levels of complexity as listed in Table~\ref{tab:model-config}. All models share a consistent input-output format and are compatible with the same audio preprocessing pipeline,
ensuring fair performance comparison across settings.

In the homogeneous setting, all clients employ the same model architecture, namely \textit{CRNN-Base}, which consists of two 1D convolutional layers (64 filters) followed by ReLU activation, max-pooling, and dropout for feature extraction, and a bidirectional GRU with 128 hidden units for temporal modeling. The extracted features are fed into a fully connected output layer for final prediction. To enable mutual knowledge transfer, each client is also equipped with a lightweight \textit{Plug-in} model, instantiated as \textit{CRNN-Lite}, which adopts a compact architecture with two 1D convolutional layers (32 filters) and a unidirectional GRU with 64 hidden units.

In the heterogeneous setting, to emulate diverse client capabilities, each client is randomly assigned one of the five CRNN variants listed in Table~\ref{tab:model-config}. These models differ in convolutional depth and GRU configuration but maintain similar classification capacity, enabling controlled architectural heterogeneity across federated clients.

\begin{table}[t]
\setlength{\tabcolsep}{3pt} 
\fontsize{8}{10}\selectfont
\centering
\caption{CRNN model variants used in homogeneous and heterogeneous client settings.}
\label{tab:model-config}
\begin{tabular}{cccc}
\hline
\textbf{Model Name} & \textbf{Conv Layers} & \textbf{RNN Type} & \textbf{Hidden Units} \\ \hline
CRNN-Tiny     & 1 × Conv1D (16 filters)       & GRU      & 32  \\ 
CRNN-Lite     & 2 × Conv1D (32 filters)       & GRU      & 64  \\ 
CRNN-Mid      & 3 × Conv1D (32 filters)       & GRU      & 64  \\ 
CRNN-Base     & 2 × Conv1D (64 filters)       & Bi-GRU   & 128 \\ 
CRNN-Deep     & 3 × Conv1D (64→128 filters)   & Bi-GRU   & 128 \\ \hline
\end{tabular}
\end{table}

\subsection{Evaluation and Implementation Details}


\textbf{Baselines.} 
We compare our method with three widely used FL algorithms: FedAvg~\cite{fedavg}, FedProx~\cite{li2020FedProx}, and FedOPT~\cite{reddi2020adaptive}. These are strong general-purpose baselines commonly applied across diverse FL tasks, including AC. We also include FedKAD~\cite{federatedtraining2023}, a recent KD-based approach designed to handle both data and model heterogeneity in FedAC. To ensure a fair comparison with our FedMLAC framework, we modify FedKAD by removing its reliance on public datasets. Specifically, we disable the use of global logits during knowledge distillation and restrict communication to class-wise aggregated feature maps only. For other FedAC methods listed in Table~\ref{tab:fAC_comparison}, fair comparison is not feasible due to the lack of standardized implementations and their dependence on incompatible datasets or architectures.


\textbf{Metrics.} We evaluate performance on the GSC dataset using classification accuracy, averaged over five independent runs with different random seeds. For the other three datasets, we adopt the F1 score as the evaluation metric, reporting the mean and standard deviation across all runs. F1 score is used to account for class imbalance, which is common in emotion recognition and sound event classification tasks.

\textbf{Hyperparameters.} In our FedAC experiments, we adopt the following training configurations to balance convergence speed and communication efficiency. The maximum number of communication rounds is set to 5000. Each client performs one local epoch per round with a batch size of 16. The learning rate is fixed at 0.01. Unless otherwise specified, 20\% of clients are randomly selected to participate in each communication round.

\subsection{Performance Under Homogeneous Models} \label{sec:performance-homo}

To assess the effectiveness of FedMLAC under homogeneous model conditions, we evaluate its performance against four representative FL baselines: FedAvg, FedProx, FedOPT, and FedKAD. In each communication round, only 20\% of clients are randomly selected to participate, except for the IEMOCAP dataset, where all clients are activated due to its small client population. As shown in Table~\ref{tab:performance-homo}, FedMLAC consistently outperforms all baseline methods across datasets and varying degrees of data heterogeneity. On GSC, a large-scale and relatively balanced dataset, FedMLAC surpasses FedOPT by 1.28\% and improves upon FedKAD by 1.96\%, demonstrating its ability to effectively integrate global knowledge even under limited client participation. Notably, while FedKAD shows stronger performance than FedAvg and FedProx, it still lags behind FedMLAC, suggesting that its distillation-based alignment alone is insufficient in capturing global generalization. For IEMOCAP, which poses challenges due to high inter-speaker variability and data scarcity, FedMLAC yields up to a 3.43\% improvement over FedAvg and 3.08\% over FedProx, and outperforms FedKAD by 1.49\%, highlighting its robustness in low-resource, high-variance environments.


\begin{table}[h]
\setlength{\tabcolsep}{1pt}
\fontsize{8}{8}\selectfont
    \centering
    \caption{Performance comparison of different FL methods (Active Ratio = 20\% except IEMOCAP) under homogeneous models.}
    \begin{tabular}{c |c| c| c| c|  c| c}
        \hline
        \textbf{Dataset} & \textbf{\begin{tabular}[c]{@{}c@{}}\#Clients \\ (Active Ratio)\end{tabular}}  & \textbf{FedAvg \cite{fedavg}} & \textbf{FedProx \cite{li2020FedProx}} & \textbf{FedOPT \cite{reddi2020adaptive}} & \textbf{FedKAD \cite{federatedtraining2023}} & \textbf{FedMLAC} \\
        \hline
        GSC  & 524 (20\%) & 80.19$\pm$0.53 & 80.43$\pm$0.47 & 83.31$\pm$0.24 & 82.63$\pm$0.39 & \textbf{84.59$\pm$0.19} \\
        \hline
        IEMOCAP & 10 (100\%) & 48.54$\pm$6.59 & 48.89$\pm$6.39 & 50.01$\pm$6.77 & 50.48$\pm$5.35 & \textbf{51.97$\pm$5.81} \\
        \hline
        CREMA-D & 18 (20\%) & 60.84$\pm$3.89 & 61.02$\pm$3.71 & 58.47$\pm$4.34 & 62.17$\pm$4.08 & \textbf{63.04$\pm$3.42} \\
        \hline
        \begin{tabular}[c]{@{}c@{}}Urban Sound \\ $\alpha=0.1$ \end{tabular} & 10 (20\%) & 50.43$\pm$4.73 & 50.62$\pm$5.58 & 48.92$\pm$5.83 & 51.92$\pm$5.49 & \textbf{52.86$\pm$4.06} \\
        \begin{tabular}[c]{@{}c@{}}Urban Sound \\ $\alpha=0.3$ \end{tabular} & 10 (20\%) & 51.83$\pm$4.42 & 52.07$\pm$4.34 & 50.36$\pm$5.61 & 52.46$\pm$4.53 & \textbf{54.29$\pm$4.17} \\
        \begin{tabular}[c]{@{}c@{}}Urban Sound \\ $\alpha=0.5$ \end{tabular} & 10 (20\%) & 53.43$\pm$4.22 & 53.61$\pm$4.01 & 51.77$\pm$5.47 & 53.88$\pm$5.39 & \textbf{55.61$\pm$3.83} \\
        \begin{tabular}[c]{@{}c@{}}Urban Sound \\ $\alpha=1.0$ \end{tabular} & 10 (20\%) & 54.64$\pm$4.40 & 54.89$\pm$3.54 & 53.03$\pm$4.93 & 54.95$\pm$5.10 & \textbf{56.95$\pm$3.65} \\
        \hline
    \end{tabular}
    \label{tab:performance-homo}
\end{table}



On the CREMA-D dataset, despite reduced client activation, FedMLAC achieves superior performance, reflecting its resilience to partial participation and speaker-specific skew. Compared to FedKAD, which already outperforms FedOPT and FedProx on this task, FedMLAC still obtains a 0.87\% gain, demonstrating the advantage of its mutual learning design. In the Urban Sound experiments, where the Dirichlet parameter $\alpha$ controls the degree of heterogeneity of federated data distributions, all methods show improved accuracy as $\alpha$ increases. However, FedMLAC consistently maintains a clear advantage across all heterogeneity levels, with accuracy rising from 52.86\% at $\alpha=0.1$ to 56.95\% at $\alpha=1.0$. Although FedKAD also shows competitive performance and benefits from increased $\alpha$, it is consistently outperformed by FedMLAC with margins ranging from 0.94\% to 2.01\%. These results collectively underscore the effectiveness of FedMLAC’s mutual learning mechanism and layer-wise robust aggregation in handling both personalization and non-IID challenges in FedAC.

\subsection{Performance Under Heterogeneous Models}

In this subsection, we assess the performance of FedKAD and FedMLAC in scenarios where clients employ heterogeneous local model architectures, which is a realistic setting in edge computing and cross-device FL. Since conventional FL baselines such as FedAvg, FedProx, and FedOPT require model consistency across clients, they are excluded from this comparison. Instead, we evaluate FedMLAC against two alternatives: FedKAD, a recent distillation-based approach designed for model heterogeneity, and an ablated variant of our method (FedMLAC w/o LPA), where the proposed LPA is disabled. All remaining experimental conditions, including client participation ratio and training protocol, are kept consistent with those in Section~\ref{sec:performance-homo} to ensure fair comparison.


\begin{figure*}[h]
\centering
\setlength{\abovecaptionskip}{0cm}
\includegraphics[width=1.0\linewidth,scale=1.0]{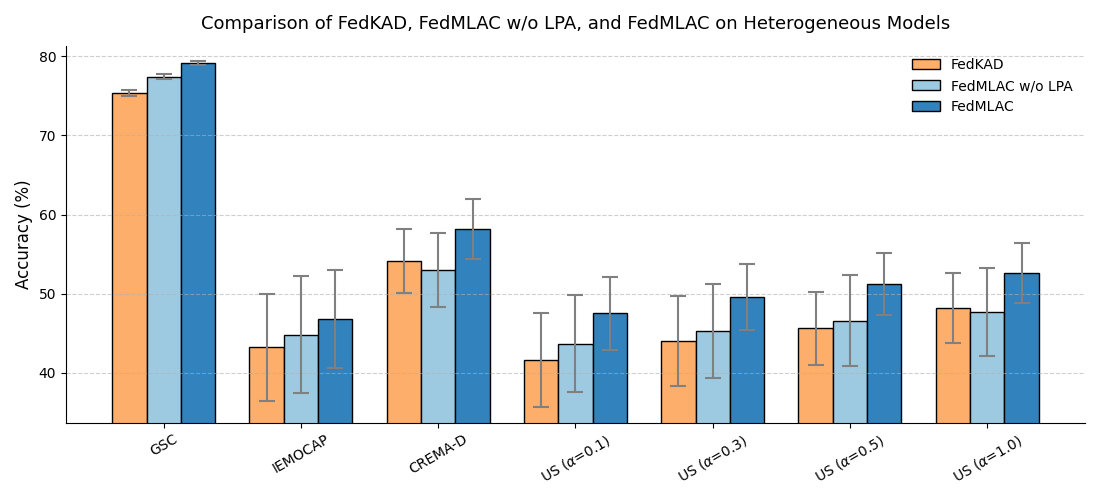}
\caption{Performance comparison among FedKAD, FedMLAC without LPA, and FedMLAC under heterogeneous model settings across four AC datasets. Each bar represents the average accuracy across all clients, with error bars indicating standard deviation. ``GSC'' denotes Google Speech Commands, and ``US'' refers to Urban Sound.}
\label{fig:hetero-performance}
\end{figure*}

As shown in Figure~\ref{fig:hetero-performance}, under heterogeneous model settings, FedMLAC consistently outperforms both its ablated variant without LPA and the KD-based baseline FedKAD across all AC datasets and non-IID configurations. Specifically, it achieves the highest accuracy across all cases, improving over its variant by up to 5.23\% on CREMA-D and over FedKAD by as much as 4.45\%. On GSC, FedMLAC reaches an accuracy of nearly 80\%, surpassing other methods with a margin of 1.75\%. For Urban Sound across different heterogeneity levels ($\alpha = 0.1$, $0.3$, and $0.5$), FedMLAC steadily leads with up to 3.82\% gains, demonstrating stable and superior performance. These results highlight FedMLAC’s robustness to model and data heterogeneity, and emphasize the critical role of LPA in filtering noisy updates and enhancing aggregation stability under challenging FedAC conditions.

\subsection{Performance Under Varying Active Ratio}

To evaluate the impact of client participation rates on model performance, we study the behavior of different FL algorithms under varying \textit{active client ratios}, ranging from 10\% to 100\%. This setting reflects real-world deployment scenarios where only a subset of clients may participate in each communication round due to resource or connectivity constraints. As client participation directly affects the data diversity seen by the server per round, it plays a critical role in model convergence and generalization. In this experiment, we exclude IEMOCAP due to its limited client pool. All other configurations, including the homogeneous model architecture and training hyperparameters, follow the setup described in Section~\ref{sec:performance-homo}.


\begin{table}[t]
\setlength{\tabcolsep}{1pt}
\fontsize{8}{10}\selectfont
    \centering
   \caption{Impact of varying active client ratios (10\%, 20\%, 50\%, and 100\%) on the performance of different FL methods under homogeneous models.}

    \begin{tabular}{c |c| c| c| c| c|  c}
        \hline
        \textbf{Dataset} & \textbf{\begin{tabular}[c]{@{}c@{}}\#Clients \\ (Active Ratio)\end{tabular}}  & \textbf{FedAvg \cite{fedavg}} & \textbf{FedProx \cite{li2020FedProx}} & \textbf{FedOPT \cite{reddi2020adaptive}} & \textbf{FedKAD \cite{federatedtraining2023}} & \textbf{FedMLAC} \\
        \hline
        \multirow{4}{*}{GSC}  & 262 (10\%) & 78.14$\pm$0.52 & 78.30$\pm$0.49 & 81.92$\pm$0.22  & 81.69$\pm$0.33 & \textbf{82.81$\pm$0.17} \\
        & 524 (20\%) & 80.19$\pm$0.53 & 80.43$\pm$0.47 & 83.31$\pm$0.24 & 82.63$\pm$0.39 & \textbf{84.59$\pm$0.19} \\
        & 1310 (50\%) & 81.73$\pm$0.58 & 81.95$\pm$0.35 & 83.82$\pm$0.31 & 83.92$\pm$0.29 & \textbf{85.10$\pm$0.21} \\
        & 2618 (100\%) & 82.58$\pm$0.46 & 82.73$\pm$0.42 & 84.84$\pm$0.19 & 84.71$\pm$0.28 & \textbf{85.76$\pm$0.13} \\
        \hline
        \multirow{4}{*}{CREMA-D} & 9 (10\%) & 59.87$\pm$4.02 & 60.10$\pm$3.78 & 58.02$\pm$4.23 & 61.14$\pm$3.65 & \textbf{61.78$\pm$3.62} \\
        & 18 (20\%) & 60.84$\pm$3.89 & 61.02$\pm$3.71 & 58.47$\pm$4.34 & 62.17$\pm$4.08 & \textbf{63.04$\pm$3.42}\\
        & 45 (50\%) & 61.72$\pm$3.75 & 61.85$\pm$4.16 & 59.52$\pm$4.05 & 62.54$\pm$3.79 & \textbf{63.93$\pm$2.24} \\
        & 91 (100\%) & 62.57$\pm$3.16 & 62.45$\pm$3.50 & 60.20$\pm$3.90 & 63.31$\pm$4.23 & \textbf{64.79$\pm$2.08} \\
        \hline
        \multirow{4}{*}{Urban Sound} & 5 (10\%) & 50.65$\pm$5.21 & 51.18$\pm$5.23 & 49.12$\pm$5.89 & 51.25$\pm$5.04 & \textbf{52.74$\pm$4.26} \\
        & 10 (20\%) & 53.43$\pm$4.22 & 53.61$\pm$4.01 & 51.77$\pm$5.47 & 53.88$\pm$5.39 & \textbf{55.61$\pm$3.83} \\
        & 25 (50\%) & 53.52$\pm$5.02 & 54.29$\pm$3.83 & 52.35$\pm$5.26 & 54.65$\pm$4.51 & \textbf{57.15$\pm$5.24} \\
        & 50 (100\%) & 55.21$\pm$3.96 & 55.47$\pm$4.74 & 53.12$\pm$4.41 & 55.97$\pm$4.87 & \textbf{58.30$\pm$3.59} \\
        \hline
    \end{tabular}
    \label{tab:performance-updated}
\end{table}


The results in Table~\ref{tab:performance-updated} show that increasing the active client ratio generally improves the accuracy of all FL methods, as more local updates contribute to global model optimization. However, FedMLAC consistently achieves the highest performance across all datasets and participation levels, and demonstrates stronger scalability with respect to client inclusion. For instance, on GSC, FedMLAC reaches 85.76\% accuracy at 100\% participation, outperforming FedOPT and FedAvg by 0.92\% and 3.18\% respectively, and exceeding FedKAD by 1.58\%. On CREMA-D, FedMLAC improves from 61.78\% at 10\% participation to 64.79\% at full participation, maintaining a consistent lead of 2–3\% over all baselines, including a 1.37\% gain over FedKAD at 100\%. Urban Sound shows a similar trend, where FedMLAC achieves 58.30\% at full participation, surpassing other methods by a margin of 2–5\%, including a 2.43\% improvement over FedKAD. These results confirm the robustness of FedMLAC in aggregating diverse local knowledge effectively, especially under partial participation.



\subsection{Performance on Noisy Audio Data}

This subsection investigates the robustness of different FL algorithms under noisy audio conditions. To simulate real-world environments with varying levels of background interference, we add white Gaussian noise to the input audio with different signal-to-noise ratio (SNR) levels. An SNR of 100\,dB corresponds to the clean, noise-free condition, while 30\,dB, 20\,dB, and 10\,dB represent increasingly severe noise levels. The Urban Sound dataset is evaluated under a moderately heterogeneous setting with $\alpha = 0.5$, which introduces realistic client-level variability alongside noise. Table~\ref{tab:performance-snr} presents the performance variations of four FL methods across all SNR conditions, with values in parentheses indicating the absolute accuracy drop from the clean condition.

\begin{table}[!t]
\setlength{\tabcolsep}{0.1pt}
\fontsize{8}{7}\selectfont
\centering
\caption{Performance comparison of different FL methods under varying SNR levels, including the clean condition (100\,dB) and decreasing levels of noise (30\,dB, 20\,dB, 10\,dB). Values in parentheses denote absolute accuracy drop from 100\,dB. Results on Urban Sound are reported under $\alpha=0.5$.}
\begin{tabular}{c |c |c| c| c| c|  c| c}
    \hline
    \textbf{Dataset} & \textbf{SNR} & \textbf{FedAvg \cite{fedavg}} & \textbf{FedProx \cite{li2020FedProx}} & \textbf{FedOPT \cite{reddi2020adaptive}} & \textbf{FedKAD \cite{federatedtraining2023}}  & \makecell{\textbf{FedMLAC} \\ \textbf{w/o LPA}} & \textbf{FedMLAC} \\
    \hline \rowcolor{gray!15}
    \multirow{8}{*}{GSC} 
    & 100dB & 80.19$\pm$0.53 & 80.43$\pm$0.47 & 83.31$\pm$0.24 & 82.63$\pm$0.39 & 82.11$\pm$0.28 & \textbf{84.59$\pm$0.19} \\
    
    & 30dB & \makecell{76.62$\pm$0.61 \\ ($\downarrow$3.57)} & \makecell{74.70$\pm$0.63 \\ ($\downarrow$5.73)} & \makecell{78.85$\pm$0.31 \\ ($\downarrow$4.46)} & \makecell{77.52$\pm$0.34 \\ ($\downarrow$5.11)} & \makecell{78.38$\pm$0.32 \\ ($\downarrow$3.73)} & \makecell{\textbf{81.75$\pm$0.16} \\ ($\downarrow$2.84)} \\
    
    & 20dB & \makecell{71.24$\pm$0.58 \\ ($\downarrow$8.95)} & \makecell{68.39$\pm$0.55 \\ ($\downarrow$12.04)} & \makecell{74.44$\pm$0.34 \\ ($\downarrow$8.87)} & \makecell{75.03$\pm$0.42 \\ ($\downarrow$7.60)} & \makecell{74.72$\pm$0.35 \\ ($\downarrow$7.39)} & \makecell{\textbf{77.28$\pm$0.29} \\ ($\downarrow$7.31)} \\
    
    & 10dB & \makecell{61.26$\pm$0.51 \\ ($\downarrow$18.93)} & \makecell{56.84$\pm$0.46 \\ ($\downarrow$23.59)} & \makecell{67.01$\pm$0.27 \\ ($\downarrow$16.30)} & \makecell{69.18$\pm$0.40 \\ ($\downarrow$13.45)} & \makecell{68.59$\pm$0.37 \\ ($\downarrow$13.52)} & \makecell{\textbf{71.27$\pm$0.33} \\ ($\downarrow$13.32)} \\
    \hline \rowcolor{gray!15}
    \multirow{8}{*}{IEMOCAP} 
    
    & 100dB & 48.54$\pm$6.59 & 48.89$\pm$6.39 & 50.01$\pm$6.77 & 50.48$\pm$5.35 & 50.25$\pm$5.91 & \textbf{51.97$\pm$5.81} \\
    
    & 30dB & \makecell{46.06$\pm$6.52 \\ ($\downarrow$2.48)} & \makecell{43.89$\pm$7.13 \\ ($\downarrow$5.00)} & \makecell{47.48$\pm$6.69 \\ ($\downarrow$2.53)} & \makecell{48.61$\pm$5.29 \\ ($\downarrow$1.87)} & \makecell{48.05$\pm$5.24 \\ ($\downarrow$2.20)} & \makecell{\textbf{50.19$\pm$5.91} \\ ($\downarrow$1.78)} \\
    
    & 20dB & \makecell{43.39$\pm$6.64 \\ ($\downarrow$5.15)} & \makecell{40.44$\pm$6.47 \\ ($\downarrow$8.45)} & \makecell{44.98$\pm$6.71 \\ ($\downarrow$5.03)} & \makecell{45.63$\pm$4.67 \\ ($\downarrow$4.85)} & \makecell{44.93$\pm$4.38 \\ ($\downarrow$5.32)} & \makecell{\textbf{48.01$\pm$4.84} \\ ($\downarrow$3.96)} \\
    
    & 10dB & \makecell{40.33$\pm$6.75 \\ ($\downarrow$8.21)} & \makecell{37.10$\pm$6.29 \\ ($\downarrow$11.79)} & \makecell{41.97$\pm$5.55 \\ ($\downarrow$8.04)} & \makecell{43.87$\pm$5.24 \\ ($\downarrow$6.61)} & \makecell{42.57$\pm$4.82 \\ ($\downarrow$7.68)} & \makecell{\textbf{45.82$\pm$6.13} \\ ($\downarrow$6.15)} \\
    \hline \rowcolor{gray!15}
    \multirow{8}{*}{CREMA-D} 
    & 100dB & 60.84$\pm$3.89 & 61.02$\pm$3.71 & 58.47$\pm$4.34 & 62.17$\pm$4.08 & 61.74$\pm$3.62 & \textbf{63.04$\pm$3.42} \\
    
    & 30dB & \makecell{58.37$\pm$3.81 \\ ($\downarrow$2.47)} & \makecell{56.01$\pm$3.83 \\ ($\downarrow$5.01)} & \makecell{55.31$\pm$5.11 \\ ($\downarrow$3.16)} & \makecell{60.31$\pm$4.62 \\ ($\downarrow$1.86)} & \makecell{60.01$\pm$3.82 \\ ($\downarrow$1.73)} & \makecell{\textbf{61.55$\pm$3.61} \\ ($\downarrow$1.49)} \\
    
    & 20dB & \makecell{54.23$\pm$4.07 \\ ($\downarrow$6.61)} & \makecell{51.91$\pm$3.96 \\ ($\downarrow$9.11)} & \makecell{50.53$\pm$4.52 \\ ($\downarrow$7.94)} & \makecell{56.48$\pm$4.18 \\ ($\downarrow$5.69)} & \makecell{56.55$\pm$4.03 \\ ($\downarrow$5.19)} & \makecell{\textbf{58.02$\pm$3.05} \\ ($\downarrow$5.02)} \\
    
    & 10dB & \makecell{50.01$\pm$3.62 \\ ($\downarrow$10.83)} & \makecell{46.23$\pm$4.08 \\ ($\downarrow$14.79)} & \makecell{47.13$\pm$4.60 \\ ($\downarrow$11.34)} & \makecell{52.83$\pm$4.16 \\ ($\downarrow$9.34)} & \makecell{51.93$\pm$4.19 \\ ($\downarrow$9.81)} & \makecell{\textbf{54.21$\pm$3.84} \\ ($\downarrow$8.83)} \\
    \hline \rowcolor{gray!15}
    \multirow{8}{*}{Urban Sound} 
    & 100dB & 53.43$\pm$4.22 & 53.61$\pm$4.01 & 51.77$\pm$5.47 & 53.88$\pm$5.39 & 54.15$\pm$3.94 & \textbf{55.61$\pm$3.83} \\
    
    & 30dB & \makecell{45.21$\pm$4.34 \\ ($\downarrow$8.22)} & \makecell{43.22$\pm$4.16 \\ ($\downarrow$10.39)} & \makecell{45.31$\pm$6.10 \\ ($\downarrow$6.46)} & \makecell{49.23$\pm$5.15 \\ ($\downarrow$4.65)} & \makecell{48.86$\pm$4.57 \\ ($\downarrow$5.29)} & \makecell{\textbf{51.23$\pm$3.96} \\ ($\downarrow$4.38)} \\
    
    & 20dB & \makecell{40.02$\pm$4.46 \\ ($\downarrow$13.41)} & \makecell{37.01$\pm$4.32 \\ ($\downarrow$16.60)} & \makecell{40.23$\pm$5.56 \\ ($\downarrow$11.54)} & \makecell{45.73$\pm$4.86 \\ ($\downarrow$8.15)} & \makecell{43.87$\pm$4.16 \\ ($\downarrow$10.28)} & \makecell{\textbf{47.32$\pm$3.70} \\ ($\downarrow$8.29)} \\
    
    & 10dB & \makecell{32.01$\pm$3.66 \\ ($\downarrow$21.42)} & \makecell{28.83$\pm$4.13 \\ ($\downarrow$24.78)} & \makecell{34.51$\pm$5.69 \\ ($\downarrow$17.26)} & \makecell{39.76$\pm$5.05 \\ ($\downarrow$13.60)} & \makecell{40.28$\pm$4.28 \\ ($\downarrow$13.87)} & \makecell{\textbf{43.03$\pm$4.13} \\ ($\downarrow$12.58)} \\
    \hline
\end{tabular}
\label{tab:performance-snr}
\end{table}

Across all datasets, FedMLAC consistently outperforms baseline methods by achieving strong accuracy under clean conditions and superior robustness across noise levels. For instance, on the GSC dataset, it achieves 84.59\% at 100\,dB and maintains 81.75\% at 30\,dB, a drop of only 2.84\%, which is smaller than FedAvg (3.57\%), FedProx (5.73\%), FedOPT (4.46\%), FedKAD (5.11\%), and FedMLAC w/o LPA (3.73\%). Similar trends are observed on IEMOCAP and CREMA-D, where FedMLAC shows the least degradation at all SNR levels. At 30\,dB, it drops only 1.78\% and 1.49\%, respectively, while FedProx and FedKAD degrade by over 5\% and 2\%, and FedMLAC w/o LPA shows 3.33\% and 2.56\% drops. Even under 10\,dB, FedMLAC remains stable, with drops of 6.15\% (IEMOCAP) and 8.83\% (CREMA-D), outperforming all comparisons. The Urban Sound dataset, characterized by high environmental variability, poses greater challenges, yet FedMLAC still demonstrates notable resilience, with a 12.58\% drop at 10\,dB, significantly less than FedAvg (21.42\%), FedProx (24.78\%), FedOPT (17.26\%), FedKAD (13.60\%), and its ablated variant (14.00\%). These results highlight FedMLAC’s practical utility in maintaining performance under real-world noisy and heterogeneous federated conditions.

\subsection{Performance on Audio Data with Label Errors}

\begin{table}[!t]
\setlength{\tabcolsep}{0.1pt}
\fontsize{8}{7}\selectfont
\centering
\caption{Performance comparison of different FL methods under varying label error rates (E.R.), including the clean condition (E.R=0.0) and increasing levels of E.R. (0.1, 0.3, 0.5). Values in parentheses denote absolute performance drop from E.R=0.0. Results on Urban Sound are reported under $\alpha=0.5$.}
\begin{tabular}{c |c |c| c| c|  c| c| c}
    \hline
    \textbf{Dataset} & \textbf{E.R.} & \textbf{FedAvg \cite{fedavg}} & \textbf{FedProx \cite{li2020FedProx}} & \textbf{FedOPT \cite{reddi2020adaptive}} & \textbf{FedKAD \cite{federatedtraining2023}} & \makecell{\textbf{FedMLAC} \\ \textbf{w/o LPA}} & \textbf{FedMLAC} \\
    \hline \rowcolor{gray!15}
    \multirow{8}{*}{GSC} 
    & 0.0 & 80.19$\pm$0.53 & 80.43$\pm$0.47 & 83.31$\pm$0.24 & 82.63$\pm$0.39 & 82.11$\pm$0.28 & \textbf{84.59$\pm$0.19} \\
    
    & 0.1 & \makecell{79.29$\pm$0.42 \\ ($\downarrow$0.90)} & \makecell{77.59$\pm$0.65 \\ ($\downarrow$2.84)} & \makecell{82.67$\pm$0.75 \\ ($\downarrow$0.64)} & \makecell{81.41$\pm$0.36 \\ ($\downarrow$1.22)} & \makecell{80.58$\pm$0.26 \\ ($\downarrow$1.53)} & \makecell{\textbf{84.07$\pm$0.68} \\ ($\downarrow$0.52)} \\
    
    & 0.3 & \makecell{74.68$\pm$0.66 \\ ($\downarrow$5.51)} & \makecell{76.12$\pm$0.41 \\ ($\downarrow$4.31)} & \makecell{78.19$\pm$0.53 \\ ($\downarrow$5.12)} & \makecell{78.92$\pm$0.41 \\ ($\downarrow$3.71)} & \makecell{78.58$\pm$0.63 \\ ($\downarrow$3.53)} & \makecell{\textbf{81.61$\pm$0.25} \\ ($\downarrow$2.98)} \\
    
    & 0.5 & \makecell{69.70$\pm$0.70 \\ ($\downarrow$10.49)} & \makecell{68.72$\pm$0.64 \\ ($\downarrow$11.71)} & \makecell{74.03$\pm$0.28 \\ ($\downarrow$9.28)} & \makecell{72.62$\pm$0.35 \\ ($\downarrow$10.01)} & \makecell{74.93$\pm$0.27 \\ ($\downarrow$7.18)} & \makecell{\textbf{77.06$\pm$0.55} \\ ($\downarrow$7.53)} \\
    \hline \rowcolor{gray!15}
    \multirow{8}{*}{IEMOCAP} 
    & 0.0 & 48.54$\pm$6.59 & 48.89$\pm$6.39 & 50.01$\pm$6.77 & 50.48$\pm$5.35 & 50.25$\pm$5.91 & \textbf{51.97$\pm$5.81} \\
    
    & 0.1 & \makecell{47.28$\pm$6.38 \\ ($\downarrow$1.26)} & \makecell{46.18$\pm$6.26 \\ ($\downarrow$2.71)} & \makecell{46.36$\pm$6.66 \\ ($\downarrow$3.65)} & \makecell{48.88$\pm$5.10 \\ ($\downarrow$1.60)} & \makecell{48.02$\pm$4.67 \\ ($\downarrow$2.23)} & \makecell{\textbf{51.13$\pm$5.37} \\ ($\downarrow$0.84)} \\
    
    & 0.3 & \makecell{43.23$\pm$6.39 \\ ($\downarrow$5.31)} & \makecell{43.36$\pm$6.26 \\ ($\downarrow$5.53)} & \makecell{40.14$\pm$6.45 \\ ($\downarrow$9.87)} & \makecell{45.83$\pm$4.87 \\ ($\downarrow$4.65)} & \makecell{45.96$\pm$5.03 \\ ($\downarrow$4.29)} & \makecell{\textbf{48.47$\pm$5.45} \\ ($\downarrow$3.23)} \\
    
    & 0.5 & \makecell{24.24$\pm$5.65 \\ ($\downarrow$24.30)} & \makecell{27.47$\pm$6.38 \\ ($\downarrow$21.42)} & \makecell{30.19$\pm$6.58 \\ ($\downarrow$19.82)} & \makecell{34.21$\pm$5.27 \\ ($\downarrow$16.27)} & \makecell{32.68$\pm$5.49 \\ ($\downarrow$17.57)} & \makecell{\textbf{36.28$\pm$5.44} \\ ($\downarrow$15.69)} \\
    \hline \rowcolor{gray!15}
    \multirow{8}{*}{CREMA-D} 
    & 0.0 & 60.84$\pm$3.89 & 61.02$\pm$3.71 & 58.47$\pm$4.34 & 62.17$\pm$4.08 & 61.74$\pm$3.62 & \textbf{63.04$\pm$3.42} \\
    
    & 0.1 & \makecell{58.23$\pm$3.59 \\ ($\downarrow$2.61)} & \makecell{59.20$\pm$3.64 \\ ($\downarrow$1.82)} & \makecell{57.75$\pm$4.55 \\ ($\downarrow$0.72)} & \makecell{60.51$\pm$4.46 \\ ($\downarrow$1.66)} & \makecell{59.06$\pm$3.63 \\ ($\downarrow$2.68)} & \makecell{\textbf{62.06$\pm$3.49} \\ ($\downarrow$0.98)} \\
    
    & 0.3 & \makecell{55.54$\pm$3.61 \\ ($\downarrow$5.30)} & \makecell{56.34$\pm$3.48 \\ ($\downarrow$4.68)} & \makecell{53.77$\pm$4.27 \\ ($\downarrow$4.70)} & \makecell{57.19$\pm$3.95 \\ ($\downarrow$4.98)} & \makecell{57.33$\pm$4.10 \\ ($\downarrow$4.41)} & \makecell{\textbf{60.58$\pm$4.28} \\ ($\downarrow$2.46)} \\
    
    & 0.5 & \makecell{47.03$\pm$3.31 \\ ($\downarrow$13.81)} & \makecell{46.40$\pm$3.64 \\ ($\downarrow$14.62)} & \makecell{48.24$\pm$4.39 \\ ($\downarrow$10.23)} & \makecell{51.66$\pm$4.32 \\ ($\downarrow$10.51)} & \makecell{52.81$\pm$3.95 \\ ($\downarrow$8.93)} & \makecell{\textbf{55.52$\pm$3.69} \\ ($\downarrow$7.52)} \\
    \hline \rowcolor{gray!15}
    \multirow{8}{*}{Urban Sound} 
    & 0.0 & 53.43$\pm$4.22 & 53.61$\pm$4.01 & 51.77$\pm$5.47 & 53.88$\pm$5.39 & 54.15$\pm$3.94 & \textbf{55.61$\pm$3.83} \\
    
    & 0.1 & \makecell{50.93$\pm$4.26 \\ ($\downarrow$2.50)} & \makecell{52.18$\pm$4.29 \\ ($\downarrow$1.43)} & \makecell{50.51$\pm$5.25 \\ ($\downarrow$1.26)} & \makecell{52.84$\pm$5.60 \\ ($\downarrow$1.04)} & \makecell{51.29$\pm$4.22 \\ ($\downarrow$2.86)} & \makecell{\textbf{54.59$\pm$3.71} \\ ($\downarrow$1.02)} \\
    
    & 0.3 & \makecell{42.66$\pm$4.31 \\ ($\downarrow$10.77)} & \makecell{47.48$\pm$4.30 \\ ($\downarrow$6.13)} & \makecell{40.19$\pm$5.63 \\ ($\downarrow$11.58)} & \makecell{47.76$\pm$5.08 \\ ($\downarrow$6.12)} & \makecell{46.82$\pm$4.62 \\ ($\downarrow$7.33)} & \makecell{\textbf{50.36$\pm$3.56} \\ ($\downarrow$5.25)} \\
    
    & 0.5 & \makecell{35.24$\pm$4.63 \\ ($\downarrow$18.19)} & \makecell{36.74$\pm$4.26 \\ ($\downarrow$16.87)} & \makecell{37.79$\pm$5.51 \\ ($\downarrow$13.98)} & \makecell{39.14$\pm$4.96 \\ ($\downarrow$14.74)} & \makecell{39.66$\pm$4.57 \\ ($\downarrow$14.49)} & \makecell{\textbf{44.98$\pm$3.54} \\ ($\downarrow$10.63)} \\
    \hline
\end{tabular}
\label{tab:performance-labelerror}
\end{table}

This subsection aims to examine how different FL methods perform under varying levels of label noise in AC scenarios. To emulate realistic annotation imperfections, we introduce varying error rates (E.R. = 0.1, 0.3, 0.5) into the training data while keeping the test sets noise-free following \cite{zhang2023fedaudio}. Experimental settings remain consistent with earlier sections, including partial client participation (20\%) for most datasets and full participation for IEMOCAP. As shown in Table~\ref{tab:performance-labelerror}, we compare the performance degradation patterns of all methods under noisy supervision, allowing us to assess their robustness across diverse acoustic tasks.

As expected, increasing label noise consistently degrades model performance across all methods and datasets, but the extent of degradation varies significantly. FedMLAC consistently demonstrates the highest resilience to label errors. On the GSC dataset at E.R.=0.5, it limits the accuracy drop to 7.53\%, compared to 10.49\% for FedAvg, 10.01\% for FedKAD, and 8.53\% for FedMLAC w/o LPA. On IEMOCAP, where label noise has a stronger effect, FedMLAC retains 36.28\% accuracy with a 15.69\% drop, smaller than FedKAD (16.27\%), FedMLAC w/o LPA (15.91\%), and significantly better than FedAvg and FedProx, which degrade by over 24\% and 21\%, respectively. Similar robustness is observed on CREMA-D and Urban Sound. On CREMA-D, FedMLAC surpasses FedAvg by 8.49\%, FedOPT by over 7\%, and FedKAD by 4.32\%. On Urban Sound, under limited client data and non-IID settings, FedMLAC exhibits the smallest drop (10.63\%) versus 18.19\% for FedAvg and 14.74\% for FedKAD. These results confirm that FedMLAC's label perturbation-aware aggregation enhances robustness under noisy supervision, underscoring its effectiveness in practical federated settings.

\subsection{Ablation Study}

To further investigate the individual contributions of each core component in the proposed FedMLAC framework, we conduct a detailed ablation study under homogeneous model settings. Specifically, we evaluate three variants: the complete \textbf{FedMLAC}, a version \textbf{without mutual learning (w/o ML)} that excludes bidirectional knowledge transfer between the local and \textit{Plug-in} models, and a version \textbf{without LPA (w/o LPA)} that replaces the robust Layer-wise Pruning Aggregation with standard averaging. This analysis aims to disentangle the effect of global knowledge sharing via mutual learning and robustness enhancement via LPA, thereby validating the design motivation behind FedMLAC. All experimental configurations, including model architectures, dataset partitions, and participation rates, are consistent with those described in Section~\ref{sec:performance-homo}.

\begin{figure*}[t]
\centering
\setlength{\abovecaptionskip}{0cm}
\includegraphics[width=1.0\linewidth,scale=1.0]{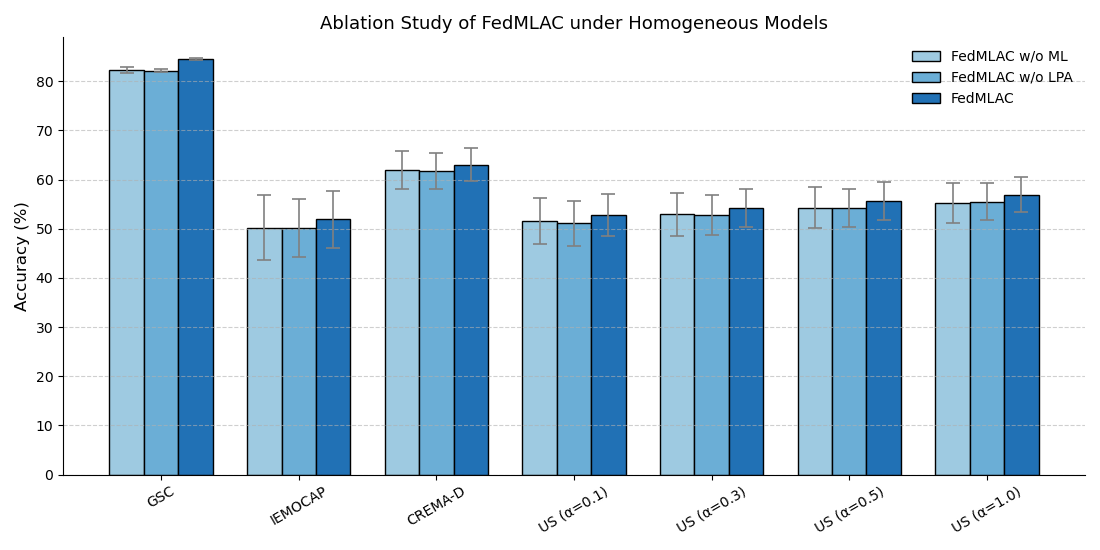}
\caption{Ablation comparison between FedMLAC and its variant without LPA and ML under heterogeneous model settings across four AC datasets. Each bar represents the average client-level accuracy, and error bars denote standard deviation across clients. ``GSC'' corresponds to Google Speech Commands, and ``US'' represents the Urban Sound dataset.}
\label{fig:ablation}
\end{figure*}

As shown in Figure~\ref{fig:ablation}, the full FedMLAC framework achieves the highest accuracy across all evaluated datasets, highlighting its strength in jointly addressing data and model heterogeneity. On GSC, removing mutual learning leads to a noticeable 2.23\% performance drop, indicating that the absence of global knowledge transfer hampers generalization. Similarly, disabling LPA results in lower accuracy compared to the complete version, especially on CREMA-D and IEMOCAP, where data noise and variability are pronounced. On Urban Sound, the full FedMLAC consistently outperforms both variants. The performance gap widens as heterogeneity increases (lower $\alpha$), demonstrating that LPA effectively mitigates biased local updates. These results confirm that both the mutual learning mechanism and the LPA module are essential for maintaining stability and boosting performance in FedMLAC



\section{Conclusion and Future Work} \label{sec:conclusion}

This work presents \textbf{FedMLAC}, a unified federated audio classification framework designed to jointly address three central challenges in real-world federated learning (FL): data heterogeneity, model heterogeneity, and data poisoning. By introducing a lightweight, globally shared \textit{Plug-in} model and a mutual learning mechanism between local and global models, FedMLAC decouples personalization from aggregation, facilitating effective cross-client knowledge transfer while supporting architectural flexibility. Moreover, to counteract the destabilizing effects of noisy or adversarial updates, a novel Layer-wise Pruning Aggregation (LPA) strategy is proposed, enhancing the resilience and stability of the global \textit{Plug-in} model. Empirical evaluations across four diverse audio benchmarks confirm that FedMLAC consistently outperforms existing FL baselines under clean and noisy conditions, both in homogeneous and heterogeneous model settings.

Despite these advancements, several open challenges remain before practical deployment. Ensuring efficient personalization in the presence of extreme client variability, minimizing communication overhead for low-resource edge devices, and maintaining robustness under adversarial manipulation or privacy-preserving constraints (e.g., differential privacy, secure aggregation) require further exploration. Addressing these general limitations will be critical for building scalable and trustworthy FL systems for audio applications in increasingly complex, decentralized environments.








\bibliographystyle{elsarticle-num} 
\bibliography{references}

\end{document}